\documentclass[%
 reprint,
superscriptaddress,
 amsmath,amssymb,
 aps,
prb,
]{revtex4-2}

\usepackage{float}
\usepackage{graphicx}
\usepackage{dcolumn}
\usepackage{bm}
\usepackage{amsmath}
\usepackage{hyperref}
\usepackage{xcolor}
\usepackage{soul}
\usepackage[justification=raggedright,singlelinecheck=false]{caption}
\usepackage{subcaption}
\begin{document}

\preprint{APS/123-QED}

\title{Shot noise signatures identifying non-Abelian properties of Jackiw-Rebbi zero modes}

\author{Haoran Ge}
\thanks{These authors equally contribute to this article.}
\affiliation{International Center for Quantum Materials, School of Physics, Peking University, Beijing 100871, China}

\author{Zhen Chen}
\thanks{These authors equally contribute to this article.}
\affiliation{International Center for Quantum Materials, School of Physics, Peking University, Beijing 100871, China}

\author{Yijia Wu}
\thanks{Corresponding author: yijiawu@fudan.edu.cn}
\affiliation{Interdisciplinary Center for Theoretical Physics and Information Sciences, Fudan University, Shanghai 200433, China}
\affiliation{State Key Laboratory of Surface Physics, Fudan University, Shanghai 200433, China}
\affiliation{Hefei National Laboratory, Hefei 230088, China}

\author{X. C. Xie}
\thanks{Corresponding author: xcxie@pku.edu.cn}
\affiliation{International Center for Quantum Materials, School of Physics, Peking University, Beijing 100871, China}
\affiliation{Interdisciplinary Center for Theoretical Physics and Information Sciences, Fudan University, Shanghai 200433, China}
\affiliation{Hefei National Laboratory, Hefei 230088, China}

\date{\today}

\begin{abstract}
Jackiw-Rebbi zero modes were first proposed in 1976 as topologically protected zero-energy states localized at domain walls in one-dimensional Dirac systems. They have attracted widespread attention in the field of topological quantum computing, as they serve as non-superconducting analogs of Majorana zero modes and support non-Abelian statistics in topological insulator systems. 
However, compared to their Majorana cousins, the braiding properties of Jackiw-Rebbi zero modes are vulnerable to the on-site energy deviation between the modes involved in the experiment. In this work, we propose to estimate the braiding properties of Jackiw-Rebbi zero-modes through measurements of transport signatures, which are readily measurable in current experiments. We find that the fidelity of braiding operation reaches unity when the current noise is fully suppressed, while this braiding fidelity monotonously decreases with the increasing of the current noise. Based on these transport signatures, we further discuss the correspondence between Majorana and Jackiw-Rebbi zero modes, highlighting their similarity in supporting non-Abelian statistics.
\end{abstract}

\maketitle

\section{\label{sec:level1}Introduction}
\par Non-Abelian anyons \cite{wilczek1982quantum} have attracted widespread attention due to their potential to provide a natural platform for fault-tolerant quantum computation \cite{nayak2008non, kitaev2003fault}. These exotic quasi-particles obey exchange statistics beyond bosons and fermions, where particle exchange performs a unitary rotation in the degenerate ground-state manifold rather than a simple phase accumulation.
Among various realizations, Majorana zero modes (MZMs) are the most studied examples of non-Abelian anyons. Originally proposed in fractional quantum Hall systems as excitations of the Moore-Read Pfaffian state at filling factor $\nu = \frac{5}{2}$ \cite{moore1991nonabelions}, MZMs were later predicted to appear in one-dimensional (1D) topological superconductors described by the Kitaev model, and to reside at the ends of a $p$-wave superconducting wire \cite{kitaev2001unpaired}. 
The non-Abelian braiding statistics of MZMs originate from the sign change $\gamma \rightarrow -\gamma$ under a $2\pi$ phase winding of the superconducting order parameter \cite{ivanov2001non}. Owing to their exceptional robustness against local perturbations, MZMs localized in half-quantum vortices are considered highly promising candidates for fault-tolerant quantum computation.

\par While significant experimental progress has been made in pursuing MZMs across diverse material platforms \cite{lutchyn2010Majorana,oreg2010Helical,mourik2012signatures,deng2016Majorana,aghaee2023Inas,aghaee2025Interferometric,wang2018Evidence,hu2024Dislocation,kong2019Half,zhu2020Nearly,zhang2018Observation,manna2020signature}, including iron-based superconductors and gold surface states, the field continues to grapple with the formidable challenge of unambiguous identification. 
The realization of MZMs demands ultra-low temperatures, atomically precise material fabrication, and sophisticated measurement techniques, making large-scale implementation prohibitively expensive and technically challenging \cite{alicea2011non,karzig2017scalable,yazdani2023Hunting}. 
Moreover, ongoing debates about distinguishing genuine MZMs from trivial bound states underscore the complexity of this endeavor.For example, in one of the most studied systems—semiconductor nanowires coupled to superconductors—the zero-bias conductance peak (ZBCP) proposed as a hallmark of MZMs has proven insufficient for definitive identification, as it can arise from various trivial mechanisms and lacks the expected correlation with topological phase transitions  \cite{mourik2012signatures,deng2012anomalous,pan2020physical,moore2018quantized,cao2023differential}. Complementary probes based on current noise have been proposed to help distinguish MZM like signals from trivial bound states such as Andreev bound states or Yu-Shiba-Rusinov states \cite{cao2023differential,k95y-7zrb}. 
Nevertheless, MZMs have so far eluded definitive experimental evidence, making the realization of non-Abelian braiding experiments using MZMs an even greater challenge.

\par Jackiw-Rebbi zero modes provide an alternative platform to realize non-Abelian anyons  \cite{klinovaja2013fractional,boross2019poor,wu2020nonabelian}. These topological zero-energy states, first proposed by Jackiw and Rebbi in 1976  \cite{jackiw1976solitons}, appear at domain walls in 1D Dirac systems with a mass term that changes sign. Numerical simulations have confirmed that Jackiw-Rebbi zero modes exhibit non-Abelian braiding properties analogous to MZMs: specifically, a Jackiw-Rebbi zero mode $\varphi$ changes its sign ($\varphi \rightarrow -\varphi$) when it encircles another Jackiw-Rebbi zero mode  \cite{wu2020double}, enabling topologically protected quantum gates  \cite{boross2019poor}. In condensed matter systems, Jackiw-Rebbi zero modes emerge as edge-localized states at both ends of the Su–Schrieffer–Heeger (SSH) chain, a one-dimensional dimerized lattice model exhibiting topologically nontrivial phases when the hopping amplitudes alternate  \cite{su1979solitons}. Crucially, unlike MZMs that require superconducting platforms and ultra-low temperatures, Jackiw-Rebbi zero modes can be realized in tunable semiconductor systems at higher temperatures, offering a more experimentally accessible route to non-Abelian physics. 
Jackiw–Rebbi zero modes realized on SSH chains have been implemented in a variety of controllable platforms ranging from semiconductor quantum dots and resonator-based circuits to optical lattices \cite{splitthoff2024gate_resonator, pham2022topological_dot, atala2012direct, Poli2014SelectiveEO}. These platforms enable direct tuning of the on-site potential and intercell coupling, making the SSH model and Jackiw–Rebbi zero modes versatile alternatives for exploring non-Abelian physics without superconductivity. 

\par The similarity in the braiding properties of Jackiw-Rebbi zero modes and MZMs motivates researchers to explore the intrinsic connection between them. From a theoretical perspective, a Jackiw-Rebbi zero mode can be regarded as  \cite{wu2023recent} a complex fermionic mode bound with a half-vortex  \cite{PhysRevE.87.052142,jackiw1981zero}. This half-vortex, which induces a non-Abelian geometric phase during braiding, also appears in Ivanov's model  \cite{ivanov2001non} describing the non-Abelian braiding of MZMs. This motivates researchers to relate Jackiw-Rebbi zero modes to symmetry-protected Majorana multiplets. The Majorana multiplets were earlier proposed in time-reversal (TR) invariant topological superconductors (TSCs) as Majorana Kramer pairs protected by TR symmetry \cite{liu2014non}. Furthermore, under unitary symmetry protection, the MZMs can also form symmetry-protected pairs, where the two MZMs within the same pair can be distinguished and labeled by different ``flavors''. During braiding, the unitary symmetry prohibits cross coupling between MZMs with different flavors, ensuring that each MZM interacts only with another MZM of the same flavor  \cite{hong2022unitary}. This unitary-symmetry-protected Majorana pair, as a whole, behaves as a complex fermionic mode supporting non-Abelian braiding, other than a self-conjugate Majorana one. For Jackiw-Rebbi zero modes with zero on-site energy difference, this unitary symmetry is preserved, and their braiding properties can be reproduced by combining the braiding behaviors of two independent sets of flavor-distinguished MZMs  \cite{wu2020double,hong2022unitary}.

\par However, despite these similarities, Jackiw-Rebbi zero modes differ fundamentally from MZMs in their robustness against local perturbations. For MZMs, quantum information is non-locally stored in a spatially separated pair, providing inherent topological protection against local disturbances. In contrast, Jackiw-Rebbi zero modes are complex fermionic zero localized at individual sites, making their non-Abelian statistics vulnerable to local on-site energy variations. Specifically, the on-site energy difference between Jackiw-Rebbi modes can break the unitary symmetry essential for non-Abelian braiding. When the on-site energy difference is zero, the unitary symmetry is preserved, and the non-Abelian braiding properties remain integral. Once the on-site energy difference is non-zero, braiding fidelity decreases with the increase of the on-site energy difference  \cite{wu2020double}. Importantly, this energy difference can be experimentally probed via electric transport measurements, which are far more accessible than performing the braiding operations themselves. This makes transport measurements a practical method for assessing the quality of non-Abelian braiding without the need for direct braiding experiments.
\par In this paper, we establish a direct correspondence between transport signatures and the fidelity of non-Abelian braiding. We begin in Sec. \ref{sec:fidelity} by mathematically defining the braiding fidelity and demonstrating its critical dependence on the on-site energy difference $M$. Motivated by the need for an accessible probe of this parameter, we then construct a transport model of Jackiw-Rebbi zero modes in Sec. \ref{sec:transportmodel} that connects $M$ to electron tunneling phenomena. We solve this model to derive our central result: a quantitative, analytical relationship between the shot noise Fano factor and the braiding fidelity. Then, in Sec. \ref{sec:JR_mzm_compare}, under the $M=0$ situation, we investigate tunneling current through Jackiw-Rebbi zero modes in the MZM basis. We validate this relationship with numerical simulations in Sec. \ref{sec:numerical} and, finally, in Sec. \ref{sec:conclusions}, we discuss the experimental implications of using noise measurements as a diagnostic tool for topological quantum operations.

\section{Braiding Fidelity and its Sensitivity to On-site Energy}
\label{sec:fidelity}
\par
\begin{figure}[t]
    \centering
    \subcaptionbox{}[1.0\linewidth]{%
        \includegraphics[width=0.95\linewidth]{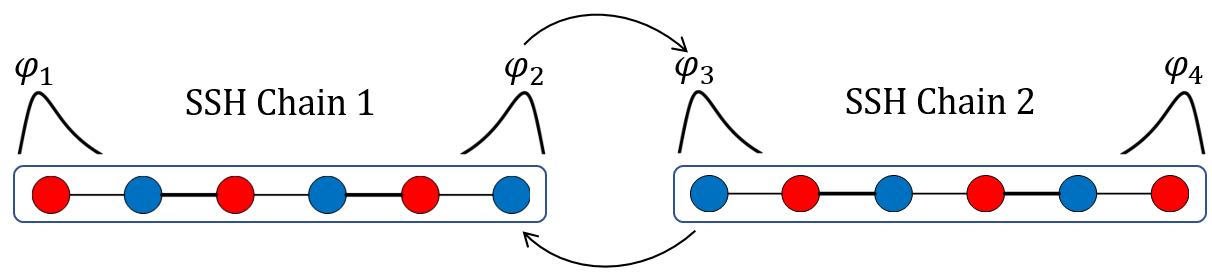}
        \label{fig:braiding_tri_junction}
    }
    \vspace{0.8em}
    \subcaptionbox{}[1.0\linewidth]{%
        \includegraphics[width=1.05\linewidth]{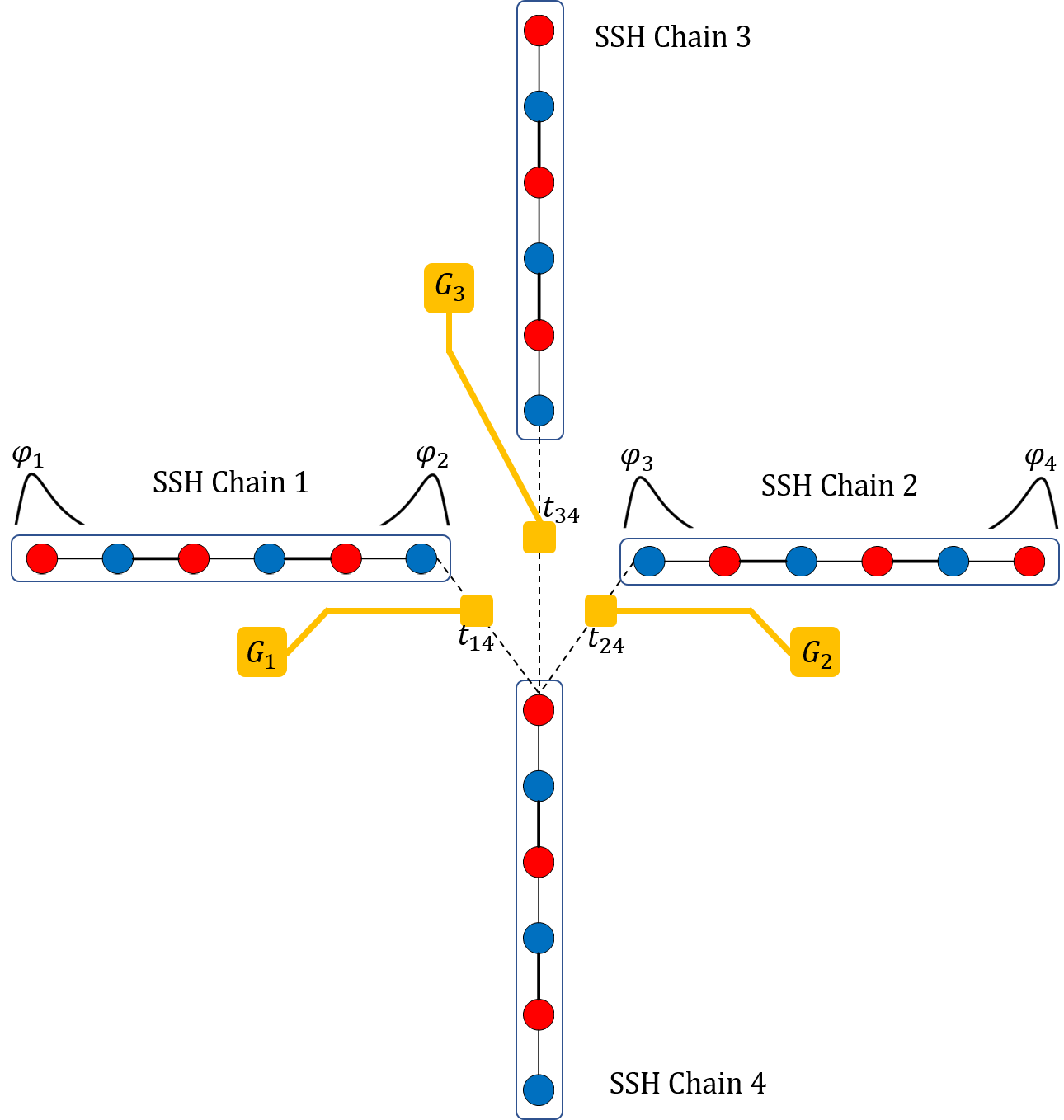}
        \label{fig:braiding_cross_junction}
    }
    \caption{(a) Two SSH chains hosting four Jackiw-Rebbi zero modes $\varphi_1$, $\varphi_2$ (in SSH chain 1) and $\varphi_3$, $\varphi_4$ (in SSH chain 2). The NOT gate is realized by swapping $\varphi_2$ and $\varphi_3$ twice in succession. (b) This braiding operation can be implemented in a cross junction via tuning tunneling amplitude $t_{14}$, $t_{24}$, $t_{34}$ between the terminal sites of different SSH chains, following Refs. \cite{amorim2015majorana,wu2020double}.}
    \label{fig:braiding schemes}
\end{figure}

 The braiding setup for Jackiw-Rebbi zero modes we investigated is based on two topologically non-trivial SSH chains [see Fig. \ref{fig:braiding schemes}(a)]. Each chain hosts two localized Jackiw-Rebbi zero modes; we denote them as $\varphi_{1}$, $\varphi_{2}$ for SSH chain 1 and $\varphi_{3}$, $\varphi_{4}$ for SSH chain 2. Take the SSH chain 1 as an example; finite size effects give rise to an overlap energy $\epsilon$ between the Jackiw-Rebbi zero modes, leading to a Hamiltonian of the form: $\epsilon \varphi^{\dagger}_{1}\varphi_{2} + \mathrm{h.c.}$. At the same time,  due to the disorder effect or gate voltages
tuning the local chemical potential, these two Jackiw-Rebbi zero modes at each end of the chain may exhibit an on-site energy difference $2M$. By choosing an appropriate zero-energy reference, this on-site energy deviation can be taken into account by adding an additional term in the Hamiltonians as: $M \varphi^{\dagger}_{1}\varphi_{1} - M \varphi^{\dagger}_{2}\varphi_{2}$.
Hence, the low-energy effective Hamiltonian describing the two Jackiw-Rebbi zero modes in the same SSH chain is written as

\begin{equation}
    H_{\mathrm{JR}} = (\epsilon \varphi_{1}^{\dagger}\varphi_{2} + \mathrm{h.c.}) + M\varphi_{1}^{\dagger}\varphi_{1} - M\varphi_{2}^{\dagger}\varphi_{2}
   \label{eq:JackiwRebbiHamiltonian}
\end{equation}

\par A basic braiding operation on this system is swapping two adjacent Jackiw-Rebbi zero modes $\varphi_{2}$ and $\varphi_{3}$ localized at different SSH chains twice in succession. Such a braiding operation can be realized via the assistance of a Y-shaped junction \cite{boross2019poor} or cross-shaped junction \cite{amorim2015majorana} [see Fig. \ref{fig:braiding schemes}(b)]. 
Swapping these two Jackiw-Rebbi zero modes once will result in $\varphi_{2} \rightarrow \varphi_{3}$ and $\varphi_{3} \rightarrow -\varphi_{2}$, which is similar to that of MZMs  \cite{wu2020double}.

\begin{figure}[t]
    \centering
    \includegraphics[width=0.99\linewidth]{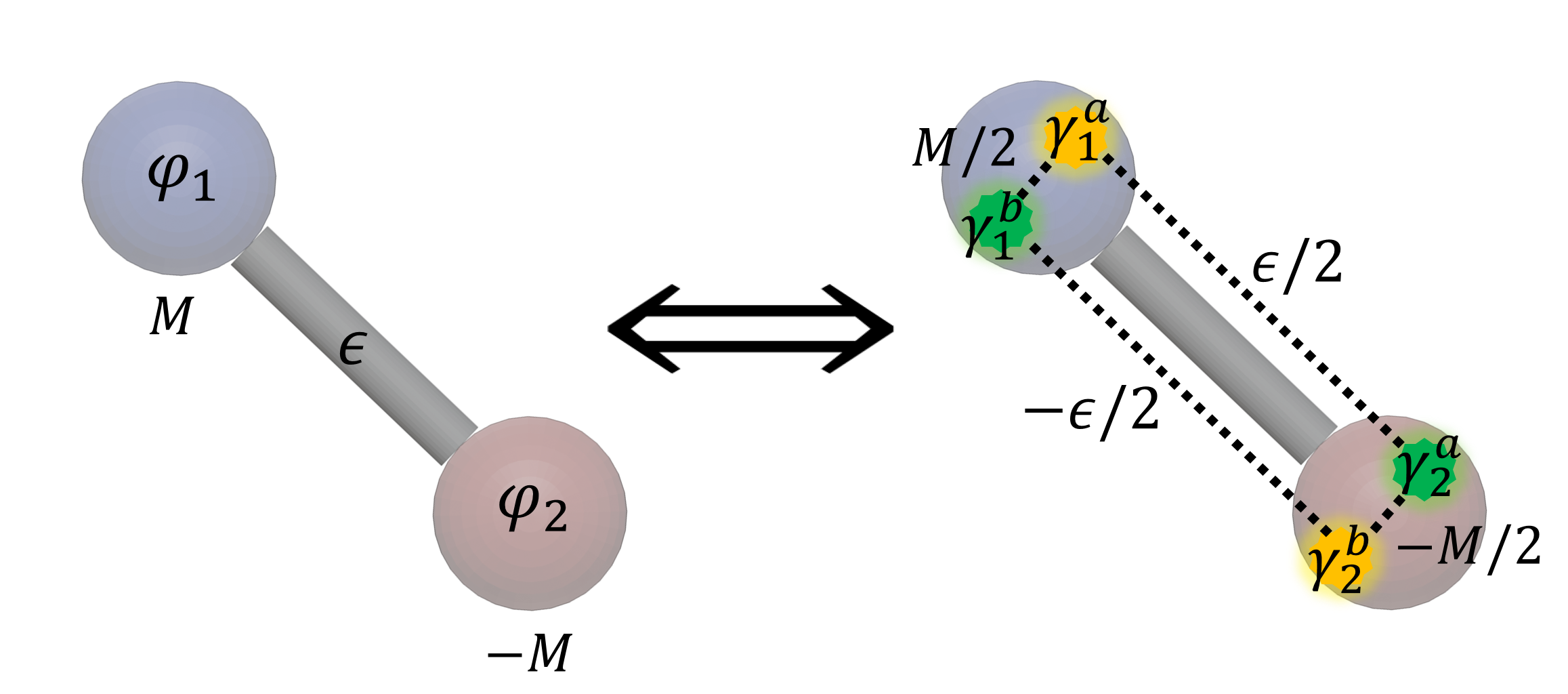}
    \caption{Decomposition of one pair of Jackiw-Rebbi zero modes into two pairs of Majorana zero modes. Left: two Jackiw-Rebbi zero modes, $\varphi_1$ and $\varphi_2$, are coupled by $\epsilon$ and have on-site energies $\pm M$, respectively. Right: the system is decomposed into Majorana components with $\varphi_1 = (\gamma_1^a + i\gamma_1^b)/2$ and $\varphi_2 = (\gamma_2^b + i\gamma_2^a)/2$. Dashed lines indicate Majorana couplings: coupling terms $\tfrac{i\epsilon}{2}\gamma_1^a\gamma_2^a$ (and $-\tfrac{i\epsilon}{2}\gamma_1^b\gamma_2^b$) between MZMs of the same ``flavor'' $a$ or $b$, and cross coupling term $\pm\tfrac{iM}{2}\gamma_j^a\gamma_j^b$ ($j=1,2$) mixing MZMs with different ``flavors''.}
    \label{fig:complex_fermion_to_mzm}
\end{figure}

\par When the on-site energy difference $M = 0$, such a braiding operation swapping these two adjacent Jackiw-Rebbi zero modes $\varphi_{2}$ and $\varphi_{3}$ twice in succession yields a negative sign for both these two modes as $|\varphi_{2} \rangle \rightarrow - |\varphi_{2} \rangle$, $|\varphi_{3} \rangle \rightarrow - |\varphi_{3}\rangle$ \cite{wu2020double}, where the quantum state is defined as $|\varphi_{i} \rangle \equiv \varphi_{i}^{\dagger} |0\rangle$. Due to the finite-size-induced coupling term above, the SSH chain considered is initially in one of its eigenstates $|\varphi^{\pm}_{12} \rangle = \frac{1}{\sqrt{2}} \left( |\varphi_1\rangle \pm |\varphi_2\rangle \right)$ before the braiding. Then the braiding operation will drive the state to evolve from $|\varphi^{\mp}_{12}\rangle$ to $|\varphi^{\pm}_{12}\rangle$. Hence, for the qubit defined based on $\left( |\varphi^{+}_{12}\rangle, |\varphi^{-}_{12}\rangle\right) ^{T}$, in the case of $M=0$, the brading operation above exchanging the two adjacent modes twice in succession is equivalent to performing a NOT gate operation on the initial state as $|\varphi_{\mathrm{final}}\rangle = X|\varphi_{\mathrm{ini}}\rangle$, where $X$ is the Pauli-X matrix. 
\par In comparison, for a non-zero on-site energy difference $M \neq 0$, although the exchange rules of the two adjacent modes remain unchanged (provided that an adiabatic condition is properly satisfied), the finite on-site energy $M$ will lead to a rotation of the eigenstates of the SSH chain  \cite{wu2020double}. Consequently, the initial state $|\varphi^{-}_{12}\rangle$ will evolve into a superposition state as $|\phi\rangle = -\sin \delta \left| \varphi_{12}^{-} \right\rangle + \cos \delta \left| \varphi_{12}^{+} \right\rangle$, other than $|\varphi^{+}_{12}\rangle$. Coefficients are given by $\cos \delta \equiv 1/\sqrt{(M/\epsilon)^{2} + 1}$, and $\sin \delta \equiv (M/\epsilon)/\sqrt{(M/\epsilon)^{2} + 1}$. In this case, to evaluate the effectiveness of the NOT gate braiding operation, we define the inner product between the target state and the final state $\mathcal{F} = |\langle \varphi_{12}^{+}|\phi\rangle|$ as a measure of fidelity. This fidelity $\mathcal{F}$ can be expressed as a function of the ratio between the on-site energy differences $M$ and the coupling energy $\epsilon$ as

\begin{equation}
    \mathcal{F} =  \frac{1}{\sqrt{(M/\epsilon)^{2} + 1}}.
\label{eq:fidelity}
\end{equation}

\par The braiding properties of Jackiw-Rebbi zero mode resemble those of MZMs when the on-site energy difference $M=0$. This can be understood from the perspective discussed in Section \ref{sec:level1} that the Jackiw-Rebbi zero mode with $M=0$ can be viewed as a unitary-symmetry-protected Majorana pair. To be specific, by decomposing the Jackiw-Rebbi zero modes into Majorana operators as $\varphi_{1} \equiv \frac{\gamma^a_1 + i\gamma^b_1}{2}$, and $\varphi_{2} \equiv \frac{\gamma^b_2 + i\gamma^a_2}{2}$, the Hamiltonian in Eq. (\ref{eq:JackiwRebbiHamiltonian}) can be written in the Majorana representation as
\begin{equation}
    H_{\mathrm{JR}} = \frac{1}{2}i\epsilon\gamma_{1}^{a}\gamma_{2}^{a} - \frac{1}{2}i\epsilon\gamma_{1}^{b}\gamma_{2}^{b} + \frac{1}{2}iM\gamma_{1}^{a}\gamma_{1}^{b} +\frac{1}{2}iM\gamma_{2}^{a}\gamma_{2}^{b}
\label{con:majora}
\end{equation}
\noindent Here the $M$ terms describe the cross coupling between the MZMs labeled by different ``flavors'' $a$ and $b$. This decomposition is schematically illustrated in Fig. \ref{fig:complex_fermion_to_mzm}. When $M = 0$, these cross coupling terms of MZMs are excluded, and this Hamiltonian actually describes the interaction between two decomposed Majorana pairs $H_{\mathrm{JR}} = \frac{1}{2}i\epsilon\gamma_{1}^{a}\gamma_{2}^{a} - \frac{1}{2}i\epsilon\gamma_{1}^{b}\gamma_{2}^{b}$. Such a Hamiltonian is protected by a unitary symmetry $\mathcal{R}$, which prohibits the cross coupling between MZMs labeled by different flavors as $\mathcal{R}i\gamma_{j}^{a}\gamma_{j}^{b}\mathcal{R}^{-1} = -i\gamma_{j}^{a}\gamma_{j}^{b}$ ($j = 1, 2$). Under such symmetry protection, a pair of Jackiw-Rebbi zero modes can be decomposed into two pairs of independent MZMs  \cite{hong2022unitary}. Therefore, braiding a pair of Jackiw-Rebbi zero modes is effectively equivalent to braiding two pairs of MZMs simultaneously, and the braiding properties of Jackiw-Rebbi zero modes proven by numerical calculation can be retrieved from the tensor product of two sets of MZMs  \cite{hong2022unitary}. As a result, Jackiw-Rebbi zero modes inherit the same braiding characteristics as MZMs. As a contrast, a non-zero on-site energy difference $M \neq 0$ will introduce cross coupling terms such as $\frac{1}{2}iM\gamma_j^a \gamma_j^b$. This indicates the Jackiw-Rebbi modes can no longer be regarded as two sets of decoupled MZMs. Consequently, the braiding properties of Jackiw-Rebbi zero modes deviate from the tensor product form, and the fidelity $\mathcal{F}$ will be lower than $1$. 

\par The analytical result in Eq. (\ref{eq:fidelity}) demonstrates that the braiding fidelity is entirely determined by the $M/\epsilon$ ratio. At the same time, directly performing the braiding operations experimentally and measuring the braiding fidelity is a challenge. This motivates our search for an alternative, accessible probe of $M$. In the following sections, we will demonstrate that electron transport signatures provide exactly such a tool.

\section{A Transport-Based Probe of Braiding Fidelity}
\label{sec:transportmodel}
\par To establish a link between the braiding fidelity, on-site energy difference, and a measurable transport signal, we now model the SSH chain as a central region connected to two metallic leads. 
The low-energy effective Hamiltonian depicting the electron tunneling mediated by these two Jackiw-Rebbi zero modes reads

\begin{equation}
    H = H'_{L1} + H'_{L2} + H_c + H_{T1} + H_{T2}
    \label{eq:TotalHamiltonian}
\end{equation}

\noindent Here, $H'_{Li} = -iv_f \int_{-\infty}^{+\infty} \mathrm{d}x' \psi_i^{\dagger}(x')\partial_x \psi_i(x')$ represents the kinetic energy of electrons propagating in the $i$-th metallic lead ($i = 1, 2$) \cite{law2009majorana}, modeled as a one-dimensional continuum with Fermi velocity $v_f$, where $\psi_i(x')$ denotes the field operator for lead $i$. $H_c = \left( \epsilon \, \varphi_1^{\dagger} \varphi_2 + \text{h.c.} \right) + M \, \varphi_1^{\dagger} \varphi_1 - M \, \varphi_2^{\dagger} \varphi_2$ is the low-energy effective Hamiltonian for the central region (the SSH chain), where $2M$ is the difference between on-site energy of two Jackiw-Rebbi zero modes. $H_{Ti} = -it_i \, \varphi_i^{\dagger} \psi_i(x = 0) + \text{h.c.}$ depicts the coupling between the $i$-th Jackiw–Rebbi zero mode and the $i$-th lead ($i = 1, 2$).
 Using the $S$-matrix approach, we can derive analytical expressions for the tunneling current and shot noise through Jackiw-Rebbi zero modes. We assume a symmetric configuration where the coupling strengths between each lead and its corresponding edge state are identical, i.e., $t_1 = t_2 \equiv t$.  The detailed derivation of the $S$-matrix elements is provided in Appendix \ref{app:smatrix}. Through our calculation, the tunneling coefficient $T(E)$ for electron transport through the Jackiw-Rebbi zero modes is given by

\begin{equation}
\label{eq:tunneling}
    T(E) = \frac{1}{(\tilde{E}_{0}^{2} - \tilde{E}^{2})^{2}/\tilde{\epsilon}^{2} + (1 + \tilde{M}^{2}/\tilde{\epsilon}^{2})}
\end{equation}

\noindent Here, we define the reduced coupling strength between the Jackiw-Rebbi zero mode and the lead as $ t_0 \equiv \frac{t^2}{v_f} $, where $v_{f}$ is the Fermi velocity. Accordingly, the reduced incoming particle energy is given by $ \tilde{E} \equiv E/t_0 $,  the reduced Jackiw-Rebbi zero mode coupling strength and the on-site energy deviation is defined as $ \tilde{\epsilon} \equiv \epsilon/t_0$, $\tilde{M} \equiv M/t_0$ and $ \tilde{E}_{0}^{2} = \tilde{\epsilon}^{2}+\tilde{M}^{2} -\frac{1}{4}$.

\begin{figure}[t]
    \centering
    \includegraphics[width=0.98\linewidth]{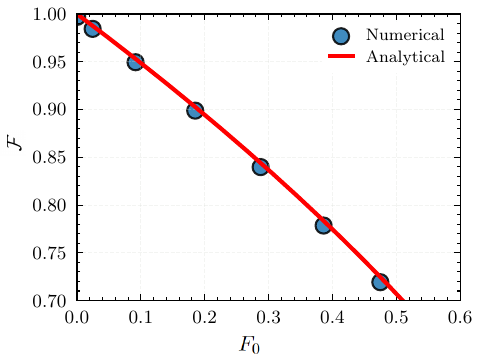}
    \caption{
    Numerical verification of the quantitative relation between the NOT gate (implemented by the braiding operations on the Jackiw-Rebbi zero modes) fidelity $\mathcal{F}$ and the Fano factor at resonant energy $F_{0}$. 
    The braiding simulation is carried out on the SSH cross junction shown in Fig. \ref{fig:braiding schemes}(b). The Hamiltonian parameters are chosen such that the finite-size coupling between Jackiw-Rebbi zero modes are $\epsilon = 6.2 \times 10^{-5}$, with on-site energy difference $M$ added at comparable energy scales. 
    Shot noise is calculated across the SSH chain 1 [see Fig. \ref{fig:braiding schemes}(b)] that the electrons are transported through the Jackiw-Rebbi zero modes $\varphi_1$ and $\varphi_2$. 
    The numerical results (blue circles) verify the analytical relation (red line) $\mathcal{F} = \sqrt{1 - F_{0}}$, confirming the relationship between the NOT gate fidelity $\mathcal{F}$ and the Fano factor at resonant energy $F_{0}$.} 
    \label{fig:fidelity_vs_fano}
\end{figure}

\par Using the tunneling current formula based on the $S$-matrix, at zero temperature, the tunneling current $I_{\mathrm{JR}}$ can be obtained by $I_{\mathrm{JR}} = \frac{e}{h}\int_{E_{f}}^{E_{f} + eV} \mathrm{d}E 
 \ T(E)$ where lead 1 is biased at voltage $V$ while lead 2 is grounded. Here $E_{f}$ is the Fermi energy. In the small voltage regime that the bias $eV \ll E_{f}$, the current $I_{\mathrm{JR}} = \frac{e^{2}V}{h}T(E_{f})$ is completely determined by the tunneling coefficient at $E_{f}$. When the Fermi level $E_{f}$ coincides precisely with the energy $E_{0}$ at resonant tunneling, the current will reach its maximum value $I_{\mathrm{JR}}^{0}$, which is determined by the ratio of on-site energy deviation $M$ and the coupling energy $\epsilon$ as  $I_{\mathrm{JR}}^{0} = \frac{e^{2}V}{h}\frac{1}{1 + M^{2}/\epsilon^{2}}$.

\par Then using the shot noise formula given by $\rm{B\ddot{u}ttiker}$  \cite{buttiker1992scattering}, we obtained that the shot noise $P_{ij}$ between leads $i$ and $j$ follows the standard relationship

\begin{equation}
\label{eq:noise}
\begin{split}
    & P_{11} = P_{22} = - P_{12} = - P_{21}\\
  = & \frac{2e^{2}}{h}\int_{E_{f}}^{E_{f} + eV}\mathrm{d}E \ T(E)(1 - T(E))\\
\end{split}
\end{equation}

\noindent Under conditions of zero temperature and a small bias voltage, the strength of the noise is given by $|P| = \frac{2e^{2}V}{h} T(E_{f})(1 - T(E_f))$. Such an expression indicates that the noise can be completely suppressed ($|P| = 0$) when the tunneling coefficient $T(E_{f})$ reaches 1 at resonant energy, i.e., on-site energy difference $M = 0$. Since noise due to thermal fluctuations is proportional to temperature $k_{B}T$ and contributes negligibly relative to shot noise in the transport process at zero temperature \cite{blanter2000shot},
an observable suppression of noise to zero can serve as an experimental signature indicating that the on-site energy difference $M$ has been tuned to zero.
\par The crucial insight is that the tunneling coefficient at resonance provides a quantitative measure of the braiding fidelity. Comparing Eq. (\ref{eq:tunneling}) with Eq. (\ref{eq:fidelity}), we find that the resonant tunneling coefficient $T(E=\pm E_0) = \mathcal{F}^{2}$. Here, the resonant energies $\pm E_0$ satisfies $E_0^2 = \epsilon^2 + M^2 -\frac{1}{4}t_0^2$. This means that when the on-site energy difference $M = 0$, both the tunneling coefficient and braiding fidelity reach their maximum values of unity. Here, we use the Fano factor $F = |P|/2eI$, which characterizes the noise relative to the current. The Fano factor $F_{0}$ at the resonant level ($E = \pm E_0$) can be expressed directly in terms of the NOT gate fidelity $\mathcal{F}$ as

\begin{equation}
    F_{0} = 1 - \mathcal{F}^{2}
\label{con:fidelity_fano}
\end{equation}

\noindent Equation (\ref{con:fidelity_fano}) provides a direct, quantitative link between a readily measurable transport quantity (the Fano factor at resonant energy, $F_{0}$) and the quality of the NOT gate implemented via non-Abelian braiding operations (the NOT gate fidelity, $\mathcal{F}$). The complete suppression of noise ($F_0=0$) corresponds to a perfect braiding operation ($\mathcal{F}=1$). As $F_{0}$ increases, the braiding fidelity $\mathcal{F}$ decreases accordingly. As shown in Fig. \ref{fig:fidelity_vs_fano}, this quantitative relationship is verified through numerical simulations of the NOT gate operation on the SSH cross junction shown in Fig. \ref{fig:braiding schemes}(b).

\section{The relation between MZMs and Jackiw-Rebbi zero modes in transport process}
\label{sec:JR_mzm_compare}

\par The transport process through Jackiw-Rebbi zero modes can also be interpreted in the Majorana basis. By decomposing each Jackiw-Rebbi zero mode into two Majorana operators $\varphi_1 \equiv \frac{\gamma^a_1 + i\gamma^b_1}{2}$ and $\varphi_2 \equiv \frac{\gamma^b_2 + i\gamma^a_2}{2}$, the Hamiltonian of the central region in Eq. (\ref{eq:TotalHamiltonian}) takes the Majorana representation form of $H_{c} = \frac{\epsilon}{2} \left( i \gamma_1^a \gamma_2^a - i \gamma_1^b \gamma_2^b \right)
+ \frac{M}{2} \left( i \gamma_1^a \gamma_1^b + i \gamma_2^a \gamma_2^b \right)$, which is exactly the same as Eq. \eqref{con:majora}. The parameter $M$ governs the cross-coupling strength between the two local Majorana components of each Jackiw–Rebbi zero mode, while $\epsilon$ characterizes the coupling between the Majorana components labeled by the same ``flavor''. Meanwhile, the Hamiltonian $H_{T1(2)}$ in Eq. (\ref{eq:TotalHamiltonian}) describing the coupling between the central region and the lead 1(or 2) is written as $H_{T1(2)} = -\frac{i}{2} t_{1(2)} \gamma_{1(2)}^{a(b)} \left[ \psi_{1(2)}(x = 0) + \psi_{1(2)}^\dagger(x = 0) \right]
- \frac{i}{2} t_{1(2)} \gamma_{1(2)}^{b(a)} \left[ \frac{ \psi_{1(2)}(x = 0) - \psi_{1(2)}^\dagger(x = 0) }{i} \right]$. In these settings, the Hamiltonian splits into three parts as
\begin{equation}
H = H_a + H_b + H_{\mathrm{coupling}}
\end{equation}

\begin{figure}[t]
    \centering
    \subcaptionbox{}[1.0\linewidth]{%
        \includegraphics[width=\linewidth]{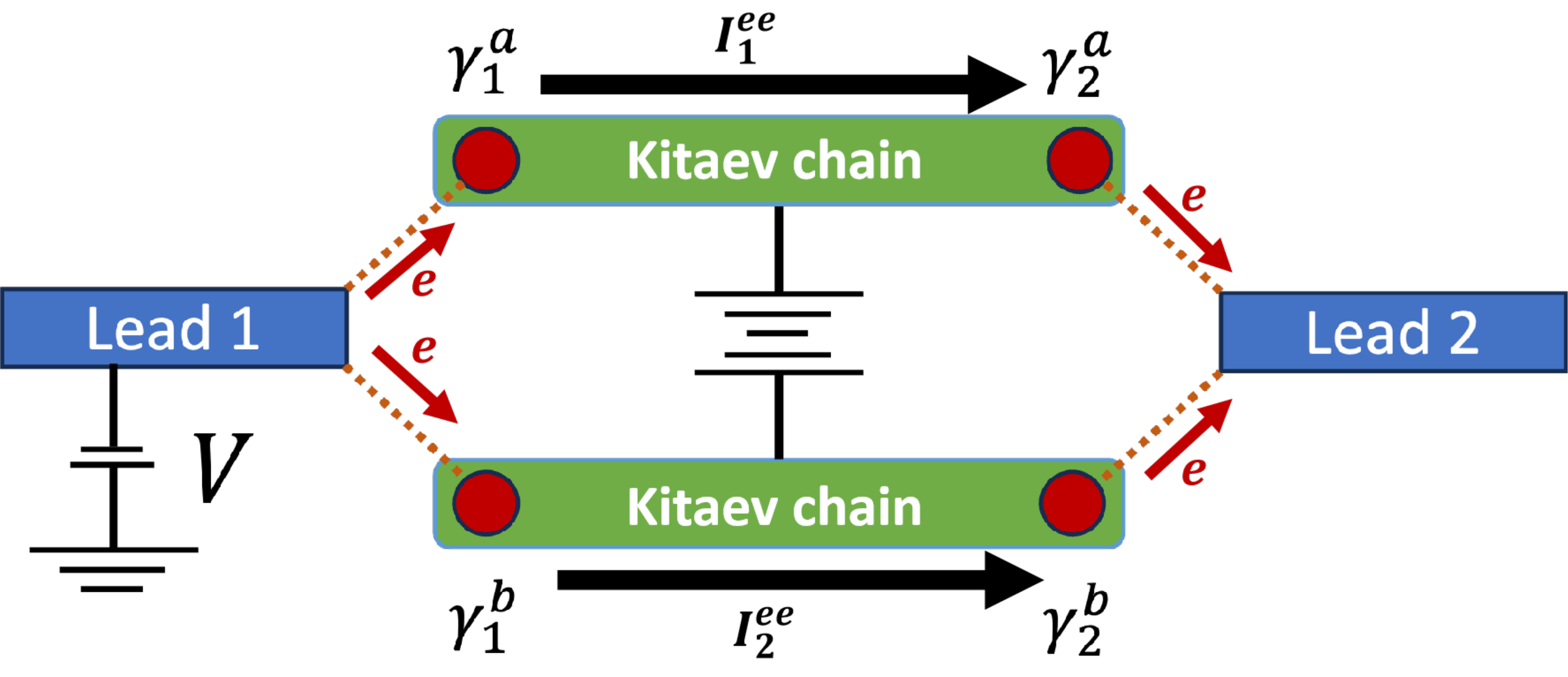}
        \label{fig:tunneling_coefficient}
    }
    \subcaptionbox{}[1.0\linewidth]{%
        \includegraphics[width=\linewidth]{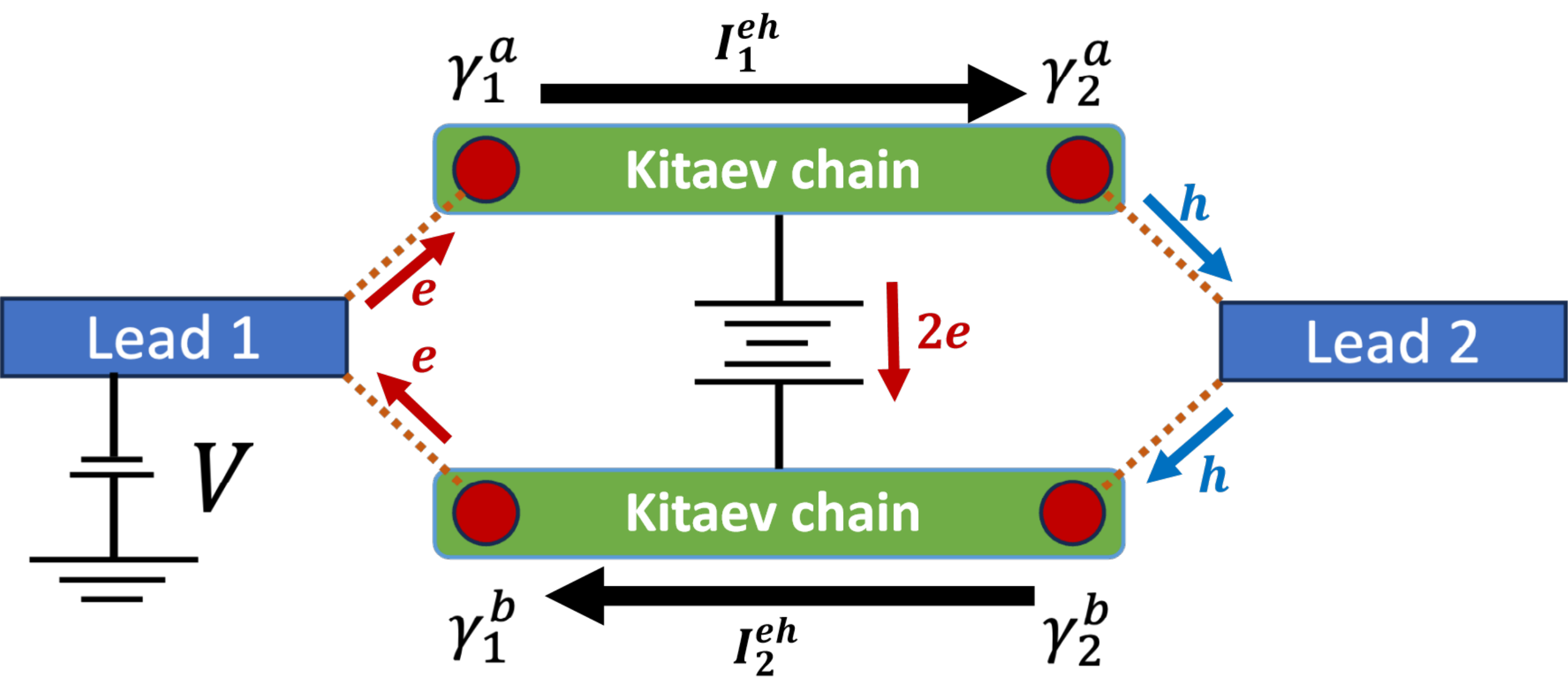}
        \label{fig:shot_noise_curves}
    }
\captionsetup{justification=raggedright,singlelinecheck=false}
    \caption{Transport through the two MZM channels $\gamma_j^a$ and $\gamma_j^b$ decomposed from a single pair of Jackiw-Rebbi zero modes. (a) The normal electron transport, where the two MZM channels act independently and each contributes to the total current. (b) The CAR process, in which charge conjugation converts the incoming electron into a hole. Due to charge conservation, a $2e$ charge Cooper pair is transferred between these two Majorana channels in the CAR, and the two MZM channels contribute the opposite current in the process, resulting in a net zero current being transmitted across the two channels.}
    \label{fig:two mzm channels}
\end{figure}

\noindent In these three parts, $H_{a} = \frac{i\epsilon}{2}\gamma_{1}^{a}\gamma_{2}^{a} -\frac{i}{2}t_{1}\gamma_{1}^{a}[\psi_{1}(0) + \psi_{1}^{\dagger}(0)]-\frac{i}{2}t_{2}\gamma_{2}^{a}[ \frac{\psi_{2}(0) - \psi_{2}^\dagger(0) }{i}]$ captures the transport through a pair of coupled MZMs $\gamma_1^a$ and $\gamma_2^a$, whereas $H_{b}= -\frac{i\epsilon}{2}\gamma_{1}^{b}\gamma_{2}^{b} -\frac{i}{2}t_{2}\gamma_{2}^{b}[\psi_{2}(0) + \psi_{2}^{\dagger}(0)]-\frac{i}{2}t_{1}\gamma_{1}^{b}[\frac{\psi_{1}(0) - \psi_{1}^\dagger(0)}{i}]$ corresponds to the transport channel via another pair of coupled MZMs $ \gamma_1^b $ and $ \gamma_2^b $. The ``cross'' coupling part $H_{\rm{coupling}}= \frac{iM}{2} (\gamma_1^a \gamma_1^b + \gamma_2^a \gamma_2^b)$ describes the local coupling between MZMs labeld by different ``flavors'' $a$ and $b$. 
In the special case $M=0$,  $H_{\rm{coupling}}$ vanishes, resulting in a complete decoupling of the Jackiw-Rebbi system into two independent Majorana channels described by $H_{a}$ and $H_{b}$, respectively.

\par By employing the $S$-matrix method (derivation details are presented in Appendix \ref{app:smatrix}), we investigated the electron transport process mediated by one of the two identical Majorana channels obtained in the decomposition above. According to the Anantram-Datta formula  \cite{anantram1996current}, the current flowing from lead 1 to lead 2 in the presence of a pair of MZMs is contributed by two parts $I_{\rm{MZM}} = I^{ee}_{\rm{MZM}} + I^{eh}_{\rm{MZM}}$. Here, at zero temperature, $I^{ee}_{\rm{MZM}} = \frac{e}{h} \int \mathrm{d}E   |S_{\mathrm{MZM}, 21}^{ee}(E)|^2 
\left\{ \theta(E_{f} - eV) - \theta(E_{f} + eV) \right\}$ describes the current contribution by normal electron transport, while $I^{eh}_{\rm{MZM}} = -\frac{e}{h}\int \mathrm{d}E |S_{\mathrm{MZM},21}^{eh}(E)|^2
\left\{ \theta(E_{f} - eV) - \theta(E_{f} + eV) \right\}$ captures the current contribution by cross-Andreev reflection (CAR) \cite{nilsson2008splitting}, a non-local electron hole transition process occurs at two leads. The $S$-matrix elements are given by $|S^{ee}_{\mathrm{MZM}, 21}(E)|^2  = |S^{eh}_{\mathrm{MZM}, 21}(E)|^2 = \frac{1}{4}T(E)$, and $T(E)$ is given by Eq. (\ref{eq:tunneling}).  

\par However, when the transport channel of Jackiw-Rebbi zero modes are decomposed into two MZMs channels, the normal electron transmission processes in the two channels remain independent [see Fig. \ref{fig:two mzm channels}(a)]. In contrast, the occurrence of CAR in the two Kitaev chains is inherently coupled, and thus the two channels cannot be treated as independent. Explicitly, the CAR in one Kitaev chain requires access to a physical ground that can supply or absorb the additional Cooper pair induced during the Andreev reflection. However, in our construction, the two Kitaev chains originate from the decomposition of a single SSH chain, which is not connected to a physical ground. As a result, the CAR processes must be accompanied by an effective Cooper-pair transfer between these two Kitaev chains.
Such a constraint implies that the electron-to-hole conversion occurring at one of the two Majorana channels is exactly reversed in the other. Consequently, the total CAR contribution from the two channels cancels out [see Fig. \ref{fig:two mzm channels}(b)]. Therefore, only the normal electron transmission contributes to the net tunneling current, and the tunneling current flowing from lead 1 to lead 2 is given by

\begin{equation}
\begin{split}
I_{\rm{MZM}}^{ee}  
= \frac{e^2 V}{h} \cdot \frac{1}{2} T(E) = \frac{1}{2}I_{\rm{JR}}^{ee}
\end{split}
\label{eq:halfcurrent}
\end{equation}

\noindent This relationship can be understood physically: when $M = 0$, the Jackiw-Rebbi system preserves a unitary symmetry that allows it to decompose into two independent Majorana transport channels. Each Majorana channel contributes exactly half the current of the original Jackiw-Rebbi system. On the contrary, a non-zero on-site energy difference $M$ will mix these two channels and alter the transport result. Remarkably, as detailed in Appendix \ref{app:SSH_Kitaev}, this correspondence manifests not only at the level of the effective Hamiltonian, where a pair of Jackiw–Rebbi zero modes can be decomposed into two pairs of Majorana zero modes, but also at the level of the lattice Hamiltonian, where an SSH chain hosting Jackiw–Rebbi zero modes can be decomposed into two independent Kitaev chains hosting MZMs \cite{verresen2017one}.
 
\section{Numerical validation for the effective model}
\label{sec:numerical}

\par To validate our analytical results based on the effective model, we perform numerical simulations using the Green's function method on lattice models of SSH and Kitaev chains. Two normal metal leads are attached to the ends of the SSH or Kitaev chains, enabling electron transport measurements through the system, and the lattice model Hamiltonian that describes the full system is written by

\begin{figure}
    \centering
    \subcaptionbox{}[1.0\linewidth]{%
        \includegraphics[width=\linewidth]{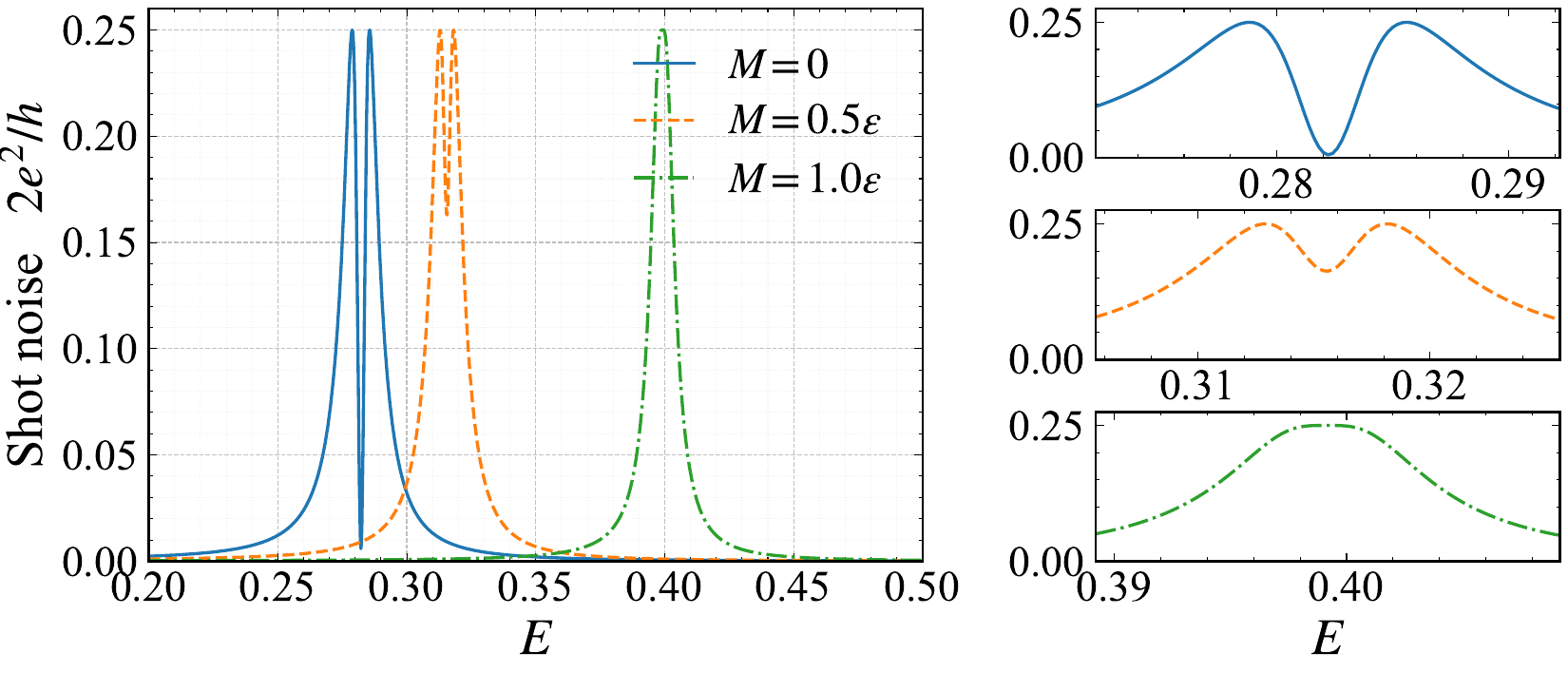}
        \label{fig:shot_noise_curves}
    }

    \vspace{0.8em}

    \subcaptionbox{}[1.0\linewidth]{%
        \includegraphics[width=\linewidth]{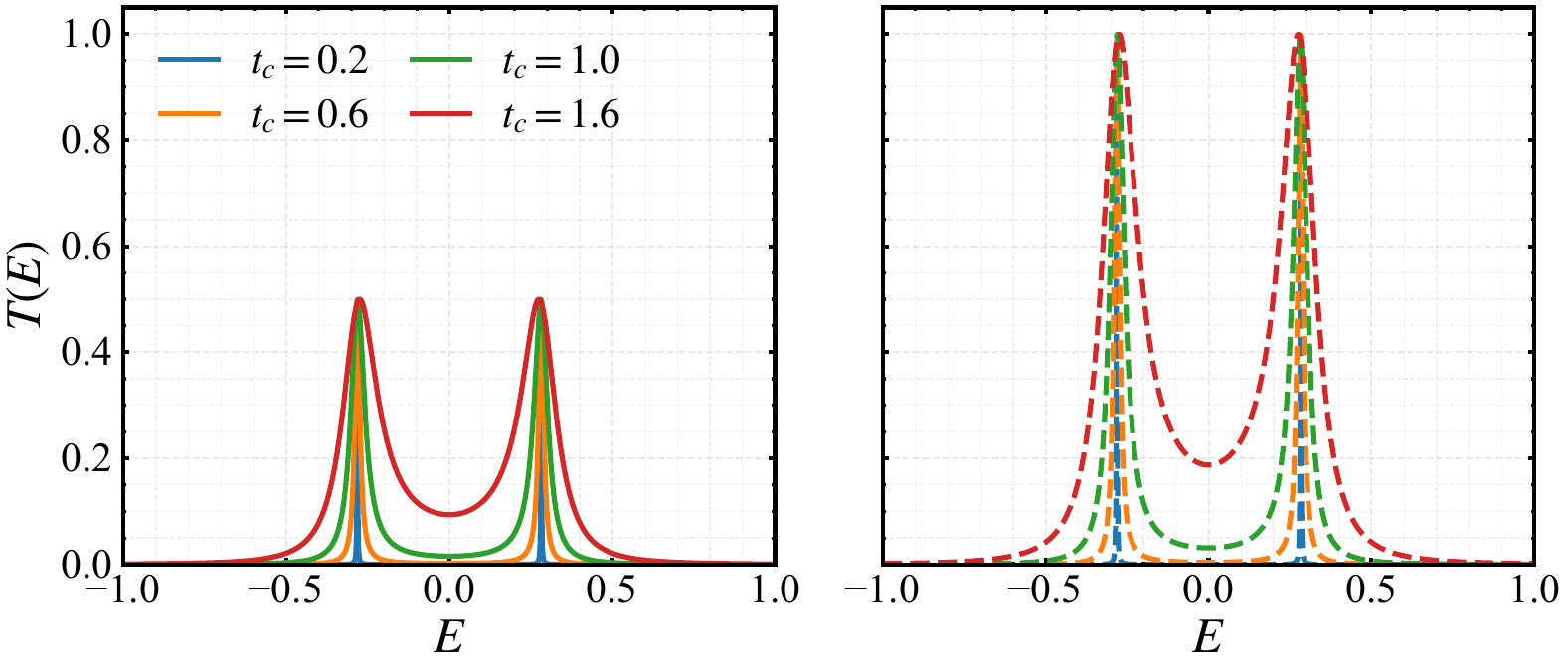}
        \label{fig:tunneling_coefficient}
    }
    \captionsetup{justification=raggedright,singlelinecheck=false}
    \caption{(a) Shot noise curves for $M = 0, 0.5\epsilon, 1.0\epsilon$ of the Jackiw-Rebbi zero modes system, showing the characteristic noise suppression at resonance. Right column: enlarged view of the noise dip. (b) Tunneling coefficient $T(E)$ as a function of energy for Kitaev chains (left) and SSH chains (right) with different coupling strengths between the central region and the lead $t_c = 0.2, 0.6, 1.0, 1.6$. The tunneling coefficient for MZMs is always half that of Jackiw-Rebbi modes as $T_{\mathrm{MZM}}(E) = \frac{1}{2}T_{\mathrm{JR}}(E)$, proving that the result shown in Eq. (\ref{eq:halfcurrent}) is independent of the coupling strength between the leads and the central region.}
    \label{fig:transport_results}
\end{figure}

\begin{equation}
    H = H_{c} + H_{L} + H_{T}
\end{equation}

\noindent Here the central region of the system described by the Hamiltonian $H_{c}$ is an SSH chain or Kitaev chain, $H_{L} = \sum_{\alpha = 1,2} \sum_{k} \epsilon_{\alpha k} a^{\dagger}_{\alpha k} a_{\alpha k}$ describes the Hamiltonian of the two leads, and $H_T = \sum_{\alpha=1,2} \sum_k \sum_i t_{i,\alpha k} c_i^\dagger a_{\alpha k} + \rm{h.c.}$ is the coupling between the central region (SSH or Kitaev chain) and two leads.
\par We simulate the transport process through SSH chains with intra-cell and inter-cell hopping amplitude $u_1 = 6$, $u_2 = 12$, and $N = 5$ unit cells (see Appendix \ref{app:greens}), giving a coupling energy $\epsilon = 0.282$ between the two Jackiw-Rebbi zero modes residing at the two ends of the chain. By tuning the on-site energy difference $M$ to values of $0$, $0.5\epsilon$, and $\epsilon$, we observe the predicted relationship between the shot noise and the on-site energy difference $M$ [see Fig. \ref{fig:transport_results}(a)]. A pronounced dip in shot noise can be observed at the resonant energy for small $M$. The shot noise at resonant energy is completely suppressed to zero when the on-site energy difference $M = 0$. In contrast, the shot noise at resonant energy increases with the increasing of the on-site energy difference $M$.

\par To compare the transport process between the MZMs and Jackiw-Rebbi modes with on-site energy difference $M=0$, we also calculate the current through a single Kitaev chain, and the parameter choice is given in Appendix \ref{app:greens}. For a single Kitaev chain decomposed from an SSH chain, the tunneling coefficient is always half that of the original SSH chain [see Fig. \ref{fig:transport_results}(b)], confirming the current relationship exhibited in Eq. (\ref{eq:halfcurrent}) and validating the Jackiw-Rebbi to Majorana correspondence in the transport process.

\section{Conclusions and discussions}
\label{sec:conclusions}
\par In summary, we have theoretically investigated the braiding fidelity and transport signatures of Jackiw-Rebbi zero modes, revealing a direct correspondence between current noise suppression and non-Abelian braiding performance. When the on-site energy difference $M = 0$, the system preserves a unitary symmetry that allows each Jackiw-Rebbi zero mode pair to decompose into two decouped MZM channels. In this regime, both the braiding and transport characteristics of Jackiw-Rebbi zero modes become equivalent to those of two independent MZM pairs, and the fidelity of the NOT gate based on the braiding operations of Jackiw-Rebbi zero modes reaches unity.
 
\par We have proposed a method for predicting NOT gate fidelity by observing transport noise that the NOT gate fidelity $\mathcal{F}$ is directly related to the Fano factor at resonant energy $F_0$ as $F_{0} = 1 - \mathcal{F}^{2}$. Below, we discuss the experimental setup designed for implementing this scheme. A representative setup is the cross-shaped junction consisting of four SSH chains [see Fig. \ref{fig:braiding schemes}(b)]. Through the operation of the central gates placed near the crossing, the Jackiw-Rebbi zero modes at the chain ends can be moved and exchanged. During the braiding process, we can connect a lead and an electric gate [not shown in Fig. \ref{fig:braiding schemes}(b)] to each end of the chain, where the electric gate is designed to manipulate the chemical potential $\mu$ of the outermost sites, and the lead is used to measure the transport current and noise. The noise between two leads connected to the same chain can serve as an observable to monitor the electric gate. We can tune the electric gate based on observed noise to ensure the on-site energy difference $M$ is zero during the braiding process.

\par At the same time, we note that the coupling energy $\epsilon$, which originates from the wavefunction overlap between different Jackiw-Rebbi zero modes, directly influences the noise spectrum. A larger $\epsilon$ corresponds to a smaller ratio $M/\epsilon$ between the on-site energy difference $M$ and the coupling energy $\epsilon$, thereby enhancing the NOT gate fidelity and yielding stronger noise suppression at the resonant energy. Hence, when selecting Jackiw-Rebbi zero modes for braiding operations, we prioritize those located in close proximity that exhibit strong coupling strength.

\begin{acknowledgments}
This work is financially supported by the National Key
R$\&$D Program of China (Grant No. 2024YFA1409000 and
No. 2019YFA0308403), the Quantum Science and Technology-National Science and Technology Major Project (Grant No. 2021ZD0302400), the National Natural Science Foundation of China (Grant No. 12304194, and No. 12574171), Shanghai Municipal Science and Technology (Grant No. 24DP2600100), Shanghai Pilot Program for Basic Research - Fudan University 21TQ1400100 (25TQ003), and Shanghai Science and Technology Innovation Action Plan (Grant No. 24LZ1400800).
\end{acknowledgments}

\appendix
\section{Equivalence between a single SSH chain and two decoupled Kitaev chains}
\label{app:SSH_Kitaev}
\par We start with an SSH chain Hamiltonian 
\begin{equation}
    H_{\rm{SSH}} = \sum_{j}(u_{1}a^{\dagger}_{j}b_{j} + u_{2}a_{j+1}^{\dagger}b_{j} + \rm{h.c.})
\end{equation}
Here $a_{j}$ and $b_{j}$ represent different sublattice sites within the $i$-th unit cell, the intra-cell hopping amplitude is $u_{1}$, and the inter-cell hopping is given by $u_{2}$. By defining two sets of Majorana operators \( \left\{\begin{array}{cc}
    \gamma_{j}^{a}\equiv a_{j}^{\dagger} + a_{j}  \\
    \gamma_{j}^{b} \equiv i (b_{j}^{\dagger} - b_{j})
\end{array} \right. \) and \( \left\{\begin{array}{cc}
    \tilde{\gamma}_{j}^{a}\equiv i(a_{j}^{\dagger} - a_{j})  \\
    \tilde{\gamma}_{j}^{b} \equiv b_{j}^{\dagger} + b_{j}  
\end{array} \right. \), the Hamiltonian of the SSH chain can be rewritten as 
\begin{equation}
\begin{split}
    H_{\rm{SSH}} &= \frac{1}{4}\sum_{j}(iu_{1}\gamma_{j}^{a}\gamma_{j}^{b} + iu_{2}\gamma_{j+1}^{a}\gamma_{j}^{b}+ \rm{h.c.}) \\
    &- \frac{1}{4}\sum_{j}(iu_{1}\tilde{\gamma}_{j}^{a}\tilde{\gamma}_{j}^{b} + iu_{2}\tilde{\gamma}_{j+1}^{a}\tilde{\gamma}_{j}^{b}+ \rm{h.c.})
\end{split}
\end{equation}
The Hamiltonian is decomposed into two parts. In each part, the decomposed Majorana Hamiltonian takes a form analogous to that of the SSH chain, where the intra-cell hopping amplitude between Majorana operators is $u_{1}$, and the inter-cell hopping is given by $u_{2}$, identical to the SSH chain. Then, by combining the Majorana operators into local complex fermion operators as \( \left\{\begin{array}{cc}
    c_{j}\equiv \frac{1}{2}(\gamma^{a}_{j} + i\gamma^{b}_{j})  \\
    c^{\dagger}_{j}\equiv \frac{1}{2}(\gamma^{a}_{j} - i\gamma^{b}_{j})
\end{array} \right. \) and \( \left\{\begin{array}{cc}
    \tilde{c}_{j}\equiv \frac{1}{2}(\tilde{\gamma}^{a}_{j} - i\tilde{\gamma}^{b}_{j})  \\
    \tilde{c}^{\dagger}_{j}\equiv \frac{1}{2}(\tilde{\gamma}^{a}_{j} + i\tilde{\gamma}^{b}_{j})
\end{array} \right. \), the SSH chain can be further written as a pair of Kitaev chains as $H_{\rm{SSH}} = H_{1}^{\rm{Kitaev}} + H_{2}^{\rm{Kitaev}}$, where 
\begin{equation}
\begin{split}
    H_{1}^{\rm{Kitaev}} &= \sum_{j}u_{1}(c_{j}^{\dagger}c_{j}-\frac{1}{2})-\sum_{j}\frac{u_{2}}{2}(c^{\dagger}_{j+1}c^{\dagger}_{j}+{\rm h.c.})\\
    &-\sum_{j}\frac{u_{2}}{2}(c_{j+1}c^{\dagger}_{j}+ \rm{h.c.})\\
    H_{2}^{\rm{Kitaev}} &= \sum_{j}u_{1}(\tilde{c}_{j}^{\dagger}\tilde{c}_{j} + \frac{1}{2})-\sum_{j}\frac{u_{2}}{2}(\tilde{c}^{\dagger}_{j+1}\tilde{c}^{\dagger}_{j}+{\rm h.c.}) \\
    &- \sum_{j}\frac{u_{2}}{2}(\tilde{c}_{j+1}\tilde{c}^{\dagger}_{j}+{\rm h.c.})
\end{split}
\end{equation}
The decomposed two Kitaev chain Hamiltonian satisfy the commutation relation that $[H_{1}^{\rm{Kitaev}}, H_{2}^{\rm{Kitaev}}] = 0$. Hence, the decomposition demonstrates that the SSH chain can be mapped onto two decoupled Kitaev chains.

\section{Detailed $S$-matrix Derivation}
\label{app:smatrix}

\subsection{$S$-matrix for Jackiw-Rebbi Zero Modes}

\par In this appendix, we provide the detailed derivation of the $S$-matrix elements for transport through Jackiw-Rebbi zero modes. The scattering matrix approach allows us to obtain analytical expressions for the tunneling current and shot noise.

\par The outgoing and incoming scattering states of electrons and holes in lead 1 (or 2) are denoted by $\psi_{1(2)E}(0^{\pm})$ and $\psi^{\dagger}_{1(2)-E}(0^{\pm})$, respectively  \cite{law2009majorana}. The $S$-matrix relates these states as $\psi_{\mathrm{out}} = S \psi_{\mathrm{in}}$, where both $\psi_{\mathrm{out}}$ and $\psi_{\mathrm{in}}$ are four-component wavefunctions defined as

\begin{equation}
\psi_{\mathrm{out}} \equiv [\psi_{1E}(0^+), \psi_{2E}(0^+), \psi^{\dagger}_{1{-}E}(0^+), -\psi^{\dagger}_{2{-}E}(0^+)]^{T}, 
\end{equation}

\noindent and
 
\begin{equation}
\psi_{\mathrm{in}} \equiv [\psi_{1E}(0^-), \psi_{2E}(0^-), \psi^{\dagger}_{1{-}E}(0^-), -\psi^{\dagger}_{2{-}E}(0^-)]^{T},
\end{equation}

\noindent respectively. Given that both electron and hole excitations are involved, the scattering matrix $S$ is represented in the electron-hole Bogoliubov-de Gennes (BdG) basis:
\begin{equation}
S = \begin{pmatrix}S^{ee} & S^{eh} \\ S^{he} & S^{hh} \end{pmatrix}
\end{equation}

\par Since Jackiw-Rebbi modes do not possess superconducting pairing terms, the off-diagonal blocks vanish: $S^{eh}_{\rm{JR}} = S^{he}_{\rm{JR}} = 0$. The electron and hole sectors are identical: $S^{ee}_{\rm{JR}} = S^{hh}_{\rm{JR}}$. For a symmetric configuration with $t_1 = t_2 \equiv t$, the scattering matrix elements of Jackiw-Rebbi zero modes are:

\begin{equation}
\begin{split}
    &S^{ee}_{\rm{JR}} = S^{hh}_{\rm{JR}} \\
  = & \frac{1}{\tilde{E}_{0}^{2} - \tilde{E}^{2} + \frac{1}{2} + i\tilde{E}}\left(\begin{array}{cc}
       \tilde{E}_{0}^{2} - \tilde{E}^{2} + i\tilde{M}  & i\tilde{\epsilon} \\
       i\tilde{\epsilon}  &  \tilde{E}_{0}^{2} - \tilde{E}^{2}- i\tilde{M} \\
    \end{array}\right)
\end{split}
\end{equation}

\noindent Here, we define the reduced parameters: $t_0 \equiv t^2/v_f$, $\tilde{E} \equiv E/t_0$, $\tilde{\epsilon} \equiv \epsilon/t_0$, $\tilde{M} \equiv M/t_0$, and $\tilde{E}_{0}^{2} = \tilde{\epsilon}^{2}+\tilde{M}^{2} -\frac{1}{4}$.

\par The tunneling coefficient is obtained from the $S$-matrix elements as

\begin{equation}
T(E) = |S^{ee}_{\rm{JR}, 21}(E)|^2 = \frac{1}{(\tilde{E}_{0}^{2} - \tilde{E}^{2})^{2}/\tilde{\epsilon}^{2} + (1 + \tilde{M}^{2}/\tilde{\epsilon}^{2})}
\end{equation}

\noindent Using the shot noise formula  \cite{buttiker1992scattering}, we obtain:
\begin{equation}
P_{ij} = \frac{2e^{2}}{h}\int_{E_{f}}^{E_{f} + eV}\mathrm{d}E \ T(E)(1 - T(E))
\end{equation}

\subsection{Connection to Majorana Zero Modes}

\par When the on-site energy difference vanishes ($M = 0$), a pair of Jackiw-Rebbi zero modes can be effectively decomposed into two independent pairs of MZMs. The electron transport mediated by one pair of MZMs can be modeled by the same Hamiltonian structure, with the Jackiw-Rebbi zero modes replaced by MZMs.

\par For a pair of MZMs coupled to the leads, the corresponding $S$-matrix is:
\begin{equation}
S_{\mathrm{MZM}} = \begin{pmatrix}I + A & A \\ A & I+A \end{pmatrix}
\end{equation}
where
\begin{equation}
A = \frac{1}{\tilde{E}_{0}^{2} - \tilde{E}^{2}+\frac{1}{2}+i\tilde{E}}\begin{pmatrix}-\frac{1}{4}-\frac{1}{2}i\tilde{E} & \frac{1}{2}i\tilde{\epsilon} \\  \frac{1}{2}i\tilde{\epsilon} & -\frac{1}{4}-\frac{1}{2}i\tilde{E} \end{pmatrix}
\end{equation}

\noindent Here $\tilde{E}_{0}^{2} = \tilde{\epsilon}^{2} -\frac{1}{4}$ because of $M = 0$. According to the Anantram-Datta formula  \cite{anantram1996current}, the current flowing from lead 1 to lead 2 in the presence of a pair of MZMs is

\begin{multline}
I_{21} = \frac{e}{h} \int \mathrm{d}E \left[ -|S_{\mathrm{MZM},21}^{ee}(E)|^2 + |S_{\mathrm{MZM},21}^{eh}(E)|^2 \right] \\ \left\{ \theta(E + eV) - \theta(E - eV) \right\}
\end{multline}

\par The transport of MZMs differs from Jackiw-Rebbi zero modes due to the Andreev reflection processes. The cross Andreev reflection  \cite{nilsson2008splitting} corresponds to the creation of a Cooper pair. At the same time, an electron from one lead is converted to a hole in the central region, then transmitted to the opposite lead. The tunneling current becomes $I_{21} \propto (-\mathcal{T} + \mathcal{T}_A)$, where $\mathcal{T} \equiv |S^{ee}_{\mathrm{MZM},21}(E)|^2$ is normal transmission and $\mathcal{T}_{A} \equiv |S^{eh}_{\mathrm{MZM},21}(E)|^2$ is the cross Andreev transformation.

\par When Jackiw-Rebbi zero modes are decomposed into two MZM channels, the normal electron transmission processes remain independent. However, as stated in Sec. \ref{sec:JR_mzm_compare} in the main text, due to charge conservation, the electron-to-hole conversion at one Majorana channel is exactly reversed at the other, causing the total cross-Andreev contribution to cancel out (see Fig. \ref{fig:two mzm channels}). Therefore, only normal electron transmission contributes:
\begin{equation}
I_{21}^{e} = \frac{2e^2 V}{h} |S_{\mathrm{MZM},21}^{ee}(E)|^2 = \frac{e^2 V}{h} \cdot \frac{1}{2} T(E)
\end{equation}

\par This result shows that the conductance associated with a pair of MZMs is exactly half that of a pair of Jackiw-Rebbi zero modes, confirming that two independent MZM channels contribute to the Jackiw-Rebbi transport.
\section{Green's Function Formalism}
\label{app:greens}

\par This appendix provides the detailed Green's function formalism used for numerical validation of our analytical results.

\par We consider a lattice model with Hamiltonian $H = H_c + H_L + H_T$, where
\begin{align}
H_c &= \text{SSH chain or Kitaev chain Hamiltonian} ,\label{eq:appendix_Hc}\\
H_L &= \sum_{\alpha = 1,2} \sum_{k} \epsilon_{\alpha k} a^{\dagger}_{\alpha k} a_{\alpha k}, \\
H_T &= \sum_{\alpha=1,2} \sum_k \sum_i t_{i,\alpha k} c_i^\dagger a_{\alpha k} + \text{h.c.}.
\end{align}

\noindent In the BdG representation, the retarded and advanced Green's functions are

\begin{equation}
G^{r/a}(E) = \frac{1}{E - H_c \pm \sum_{\alpha = 1, 2} \frac{i}{2} \Gamma^{\alpha}_{\mathrm{BdG}}} = \left(\begin{array}{cc}
G^{ee}  & G^{eh}  \\
G^{he}  & G^{hh}  \\
\end{array}\right)
\end{equation}

\noindent Here, the line-width function is

\begin{equation}
\Gamma^{\alpha}_{\mathrm{BdG}} = \left(\begin{array}{cc}
\Gamma^{\alpha}  & 0  \\
0  & (\Gamma^{\alpha})^{T}  \\
\end{array}\right)
\end{equation}

\noindent where $\Gamma^{\alpha}_{ji}(E) = 2\pi\, t^{*}_{i,\alpha k}\, t_{j,\alpha k}\, \rho(E)$ characterizes the coupling between lead $\alpha$ and the central region, 
$\rho(E)$ is the density of states in the corresponding lead.

\par The tunneling coefficient is expressed as:
\begin{equation}
T(E) = \text{Tr}\left[ \Gamma^1_{\rm{BdG}} G^r \Gamma^2_{\rm{BdG}} G^a \right]_{ee}
\end{equation}

\par For SSH chains, we introduce on-site energy differences $M$ via adding

\begin{equation}
H_{\rm{onsite}}=\sum_{j}M(a^{\dagger}_{j}a_{j}-b_{j}^{\dagger}b_{j})
\end{equation}

\noindent into $H_c$ in Eq. (\ref{eq:appendix_Hc}). The parameters used in our simulations (see Fig. \ref{fig:transport_results}) are: $u_1 = 6$, $u_2 = 12$, $N = 5$, giving $\epsilon = 0.282$. For Kitaev chains, the corresponding parameters are: $\mu = u_1$, $t_{p} = \Delta_p = u_2/2$. Here $\mu$ is the chemical potential, $t_{p}$ is the hopping amplitude, and $\Delta_{p}$ is the $p$-wave superconducting potential of the Kitaev chain. It is worth noting that the coupling strengths between the lead and the central region are the same as those of the SSH chain.

\bibliography{apssamp}

\end{document}